\documentclass[5p]{elsarticle}

\usepackage{graphicx}
\usepackage{dcolumn}
\usepackage{bm}
\usepackage{hyperref}
\usepackage[T1]{fontenc}
\usepackage[utf8]{inputenc}
\usepackage{pslatex}
\usepackage{textcomp} 
\usepackage{color}
\usepackage{lmodern} 
\usepackage[normalem]{ulem}
\usepackage{amsmath}
\usepackage{amsfonts}
\usepackage{breakurl} 
\usepackage{array,multirow,graphicx}
\usepackage{lscape}

\hypersetup{
    breaklinks = true,
    colorlinks = true,
    pdftitle = {}, 
    pdfauthor = {},
    pdfkeywords = {},
    linkcolor = blue,
    citecolor = blue,
    filecolor = black,
    urlcolor = magenta
}

\makeatletter 
\def\@author#1{\g@addto@macro\elsauthors{\normalsize%
    \def\baselinestretch{1}%
    \upshape\authorsep#1\unskip\textsuperscript{%
      \ifx\@fnmark\@empty\else\unskip\sep\@fnmark\let\sep=,\fi
      \ifx\@corref\@empty\else\unskip\sep\@corref\let\sep=,\fi
      }%
    \def\authorsep{\unskip,\space}%
    \global\let\@fnmark\@empty
    \global\let\@corref\@empty  
    \global\let\sep\@empty}%
    \@eadauthor={#1}
}
\makeatother

\graphicspath{{./figures/}}
\DeclareUnicodeCharacter{1EF3}{\`y}

\newcommand{\yfeco}{Y(Fe$_{1-x}$Co$_x$)$_2$} 
\newcommand{\yfe}{YFe$_2$} 
\newcommand{\zrfe}{ZrFe$_2$} 
\newcommand{\yco}{YCo$_2$}
\newcommand{\zrco}{ZrCo$_2$}
\newcommand{\zrfeco}{Zr(Fe$_{1-x}$Co$_x$)$_2$}

\begin{document}

\begin{sloppypar}

\title{Curie temperature study of the \yfeco{} and \zrfeco{} systems using mean-field theory and Monte Carlo method}

\author{Bartosz Wasilewski}
\address{Institute of Molecular Physics, Polish Academy of Sciences,\\ M. Smoluchowskiego 17, 60-179 Pozna\'n, Poland}

\author{Wojciech Marciniak\corref{cor1}}
\cortext[cor1]{Corresponding author} 
\ead{wojciech.ro.marciniak@student.put.poznan.pl}
\address{Institute of Physics, Faculty of Technical Physics, Poznań University of Technology, Piotrowo 3, 61-138 Pozna\'{n}, Poland}

\author{Miros\l{}aw Werwi\'nski}
\address{Institute of Molecular Physics, Polish Academy of Sciences,\\ M. Smoluchowskiego 17, 60-179 Pozna\'n, Poland}

\begin{abstract}

The cubic Laves phases including \yfe{}, \yco{}, \zrfe{}, and \zrco{} 
are considered as promising candidates for application in hydrogen 
storage and magnetic refrigeration.
While \yfe{} and \zrfe{} are ferromagnets, alloying with Co decreases 
magnetic moments and Curie temperatures ($T_\mathrm{C}$) of pseudobinary 
\zrfeco{} and \yfeco{} systems, leading to the paramagnetic states of \yco{} and \zrco{}.
The following study focus on the investigatation of Curie temperature of 
the \yfeco{} and \zrfeco{} system from first principles.
To do it the Monte Carlo (MC) simulations and the mean field theory (MFT) based on the disordered local 
moments (DLM) calculations are used.
The DLM-MFT results agree qualitatively with the experiment and 
preserve the characteristic features of
$T_\mathrm{C}(x)$ dependencies for both \yfeco{} and \zrfeco{}.
However, we have encountered complications in the Co-rich regions due to 
failure of the local density approximation (LDA) in describing the Co 
magnetic moment in the DLM state.
The analysis of Fe-Fe exchange couplings for \yfe{} and \zrfe{} phases 
indicates that
the nearest-neighbor interactions play the main role in the formation 
of $T_{\mathrm{C}}$. 

\end{abstract}

\date{\today}

\maketitle
\section{Introduction}
Laves phases are close packed structure intermetallics with a chemical composition AB$_2$.
They are classified into three types: hexagonal MgZn$_2$ (C14), cubic MgCu$_2$ (C15), and hexagonal MgNi$_2$ (C36).
In this work we investigate theoretically the C15 cubic phases \yfe{} and \zrfe{} together with their pseudobinary alloys with \yco{} and \zrco{}.
The mentioned systems were recently intensively studied from both fundamental and application points of view.
For hydrogen storage applications were considered the \zrfe{}, \yfe{} and its alloys~\cite{wiesinger_structural_2005,isnard_pressure-induced_2011,li_reversible_2016} together with Zr-Fe-Co~\cite{jat_structural_2015} and Zr(Cr$_{0.5}$Ni$_{0.5}$)$_2$~\cite{merlino_dft_2016} ternary alloys.
\zrfe{} and its role in the hydrogen storage behaviour of selected hydride systems has also been discussed~\cite{shahi_mgh2zrfe2hx_2015,shukla_enhanced_2017}.
Furthermore, \yco{} alloys with rare-earth elements R$_{1-x}$Y$_{x}$Co$_2$ (R~=~Er,~Gd) were investigated as magnetocaloric materials for application in magnetic refrigerators~\cite{baranov_butterflylike_2009,pierunek_normal_2017}, similar like Er$_{1-x}$Zr$_{x}$Fe$_2$ alloys~\cite{mican_magnetic_2013}.
Above efforts are supplemented by a number of theoretical studies of mechanical~\cite{chen_phase_2015}, electronic~\cite{bhatt_high_2016}, and magnetic properties~\cite{zhang_prediction_2015,zhuravleva_magnetovolume_2017} concerning \zrfe{}, \yco{}, and \zrco{} compounds.
The binary XFe$_2$ and XCo$_2$ phases (including X~=~Y,Zr) have been also investigated theoretically from a permanent magnets perspective~\cite{kumar_permanent_2014}.

One of the key physical quantities of a magnetic material is the Curie temperature ($T_{\mathrm{C}}$) indicating the magnetic phase transition.
It is known from experiment  that the \yfe{} and \zrfe{} are ferromagnets with Curie  temperatures  of 550 and 620~K, respectively~\cite{hilscher_low_1978, guzdek_electrical_2012}.
\yco{} is an exchange enhanced Pauli paramagnet and \zrco{} is a regular Pauli paramagnet~\cite{yamada_nmr_1980}.
Furthermore, \yco{} undergoes a  metamagnetic transition at 70~T and 10~K~\cite{goto_itinerant_1990}, while \zrco{} does not exhibit such transition~\cite{yamada_itinerant_1990}.
In \yco{} magnetic ordering can be also induced by introducing either defects or chemical disorder~\cite{sniadecki_induced_2014,sniadecki_magnetic_2015}.
Taking into account the recent interest in  \yfe{}, \zrfe{}, \yco{}, and \zrco{} we decided to investigate from first principles the $T_{\mathrm{C}}$ of the above compounds and their pseudobinary alloys.
We try to explain the dependence of $T_{\mathrm{C}}$ on first neighbor interactions for \yfe{} and \zrfe{} and to draw a relation between the $T_{\mathrm{C}}$ and chemical composition.
Out of several methods that allow to study the $T_{\mathrm{C}}$ from first principles,
we use in this work the mean-field theory based on the disordered local moment method (DLM-MFT) and Monte Carlo (MC) method. 
The MC method allows us to make predictions by simulating the Heisenberg model of a given composition, such as magnetization ($M$) and susceptibility ($\chi$) vs temperature dependencies\cite{eriksson_atomistic_2017}, while using the DLM-MFT we can calculate only the Curie temperature of a given composition.

\section{Calculations' details}

The Curie temperature can be estimated with the use of the mean-field theory (DLM-MFT) by considering the difference in energy between the DLM and ferromagnetic states\citep{gyorffy_first-principles_1985,bergqvist_theoretical_2007} according to the following equation:
\begin{equation}\label{eq:1}
 T_\mathrm{C}^{\mathrm{DLM-MFT}} = \frac{2}{3} \frac{E_{\mathrm{DLM}} - E_{\mathrm{FM}}}{k_\mathrm{B}\times c },
\end{equation}
for $E_{\mathrm{DLM}} - E_{\mathrm{FM}} > 0$, and
\begin{equation}
T_\mathrm{C}^{\mathrm{DLM-MFT}} = 0,
\end{equation}
otherwise.
The $E_{\mathrm{DLM}}$ and $E_{\mathrm{FM}}$ are respectively the total energies for the DLM and ferromagnetic states, $k_{\mathrm{B}}$ is the Boltzmann constant, and $c$ is the concentration of magnetic atoms.
In this work we use the coherent potential approximation (CPA)~\cite{soven_coherent-potential_1967} in two ways. 
One way is to treat the chemical disorder (ferromagnetic states) and the other is to treat the magnetic and chemical disorders in the DLM calculations. 
Here, we simultaneously use both methods. 
Magnetically disordered state is done by forming a model with half of the magnetic moments pointing one direction and half pointing the opposite. 
For example, to model the paramagnetic state of Zr(Fe$_{0.5}$Co$_{0.5}$)$_2$ we create a following configuration 
Zr(Fe$_{0.25\uparrow}$Fe$_{0.25\downarrow}$Co$_{0.25\uparrow}$Co$_{0.25\downarrow}$)${_2}$. 
The arrows indicate the direction of the magnetic moment on each atom. 
The magnetic moment on Zr is of induced character and there was no need to treat it with the DLM. 

The second method used in this work to determine the Curie temperature is MC simulations. 
The classical (i.e. not quantum) MC simulations were done using the Uppsala atomistic spin dynamics (UppASD) 
code~\cite{skubic_method_2008} with the magnetic moments and exchange integrals obtained from the spin polarized relativistic Korringa-Kohn-Rostoker (SPR-KKR) 
\cite{ebert_et_al._munich_nodate,ebert_calculating_2011} calculations. 
The exchange integrals were calculated using the method of Liechtenstein \textit{et al.} \cite{liechtenstein_exchange_1984} with respect to the ferromagnetic state. 
In the MC simulations we determine $T_\mathrm{C}$ from the position of the peak in the temperature dependence of susceptibility. 
The radius of the exchange integrals cutoff sphere in the Heisenberg model was set up to 1.5 lattice parameter (\textit{a}) which means that only the atomic pairs separated by distance $\leq$ 1.5~\textit{a} were considered. 
The simulated system consisted of 8800 atoms with periodic boundary conditions. 
The simulations have been checked for convergence with the radius of the Heisenberg model cutoff sphere. 
In this work the DLM calculations were performed for the whole range of Co concentrations, while the MC simulations were limited to \yfe{} and \zrfe{}.

 \begin{figure}[ht]
 \centering
\includegraphics[width=0.8\columnwidth]{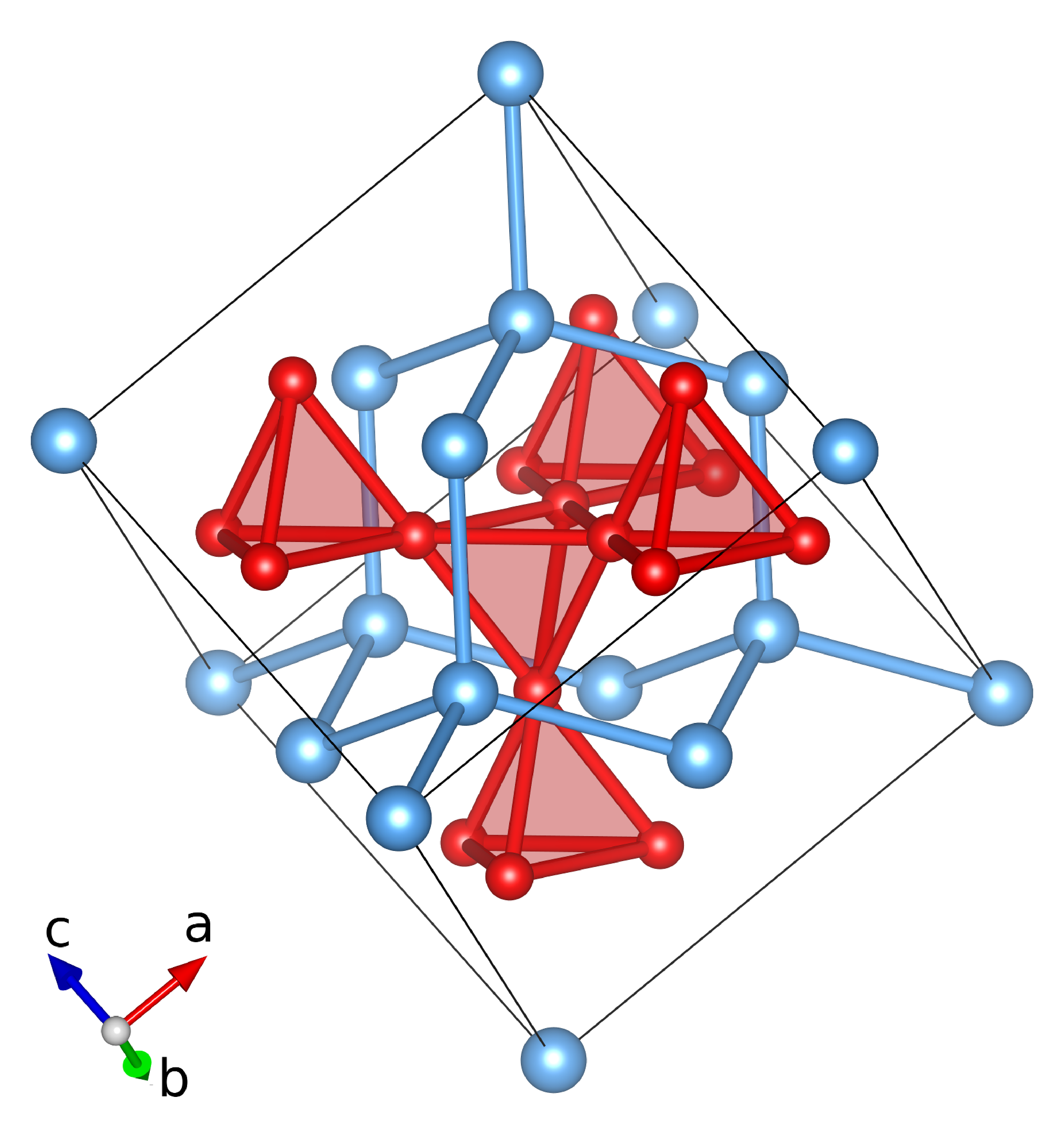}
\caption{\label{fig:struct}
Crystal structure of the cubic MgCu$_2$-type Laves phase. The large blue balls represent Mg atoms and the small red balls indicate Cu atoms.
}
\end{figure}
\begin{table}[ht]
\centering
\caption{\label{tab:wyckoff} 
Atomic coordinates for \yfeco{} and \zrfeco{}, space group $Fd$-$3m$ (no. 227), origin choice two. 
}
\vspace{1mm}
\begin{tabular}{|l|l|ccc|}
\hline \hline
atom  	& site	& $x$ 	& $y$ & $z$ \\
\hline
Y/Zr   	& 8(a) 	& 1/8 	& 1/8   & 1/8 \\	
Fe/Co  	& 16(d)	& 1/2 	& 1/2   & 1/2 \\
\hline \hline
\end{tabular}
\end{table}

The occupied atomic position and the space group of the C15 cubic Laves phase can be found in Tab. \ref{tab:wyckoff}. The unit cell is presented in Fig. \ref{fig:struct}, which was created with the use of the VESTA code\citep{momma_vesta_2011}. For the calculations of the binary compositions we used the experimental values of the lattice parameters which are equal to
7.36\citep{guzdek_electrical_2012}, 7.22\cite{guzdek_electrical_2012},  7.07\citep{muraoka_magnetic_1979}, and 6.96\cite{muraoka_magnetic_1979}~\AA{} for \yfe{}, \yco{}, 
\zrfe{}, and \zrco{} respectively. Since in the experiment the lattice parameters of \yfeco{} and \zrfeco{} change almost linearly with Co concentration\cite{muraoka_magnetic_1979,guzdek_electrical_2012}, for the intermediate composition we assumed linear behavior of the lattice parameters. 
The electronic band structure calculations for the whole concentration 
range of \yfeco{} and \zrfeco{} systems were performed using the 
full-potential local-orbital minimum-basis scheme 
FPLO5.00~\cite{koepernik_self-consistent_1997, 
koepernik_full-potential_1999}.
The chemical and magnetic disorder on the Fe/Co site was modeled with the CPA.
Due to limitations of FPLO5.00, we used the scalar-relativistic approach 
and local density approximation in the form of Perdew and Wang (PW92)~\cite{perdew_accurate_1992}.
The basis was optimized.
The 3$s$3$p$ Fe/Co orbitals were treated as semicore and 4$s$4$p$3$d$ as valence.
The Y and Zr 4$s$4$p$ orbitals were treated as semicore and 5$s$5$p$4$d$ as valence.
After convergence tests we have chosen a $12 \times 12 \times 12$  \textbf{k}-mesh.
We applied simultaneous energy and charge density convergence criteria of $\sim$2.72$\times$10$^{-7}$~eV (10$^{-8}$~Ha) and 10$^{-6}$, respectively.

The MC simulations were performed using magnetic moments and exchange couplings obtained with the version 7.6 
of the SPR-KKR code. 
The  calculations were performed using 40 energy points on a semicircular energy path and 1000 irreducible \textbf{k}-points, corresponding to a $36 \times 36 \times 36$ mesh. For the exchange-correlation potential we employed the Vosko-Wilk-Nusair (VWN) \citep{vosko_accurate_1980} form of the local density approximation (LDA). The SPR-KKR calculations were done in the full relativistic and full potential approaches.

\section{Results and Discussion}
\subsection{Y(F\lowercase{e}$_{1-x}$C\lowercase{o}$_x$)$_{2}$}

\begin{figure}[ht]
\includegraphics[trim = 0 0 0 0,clip,width=\columnwidth]{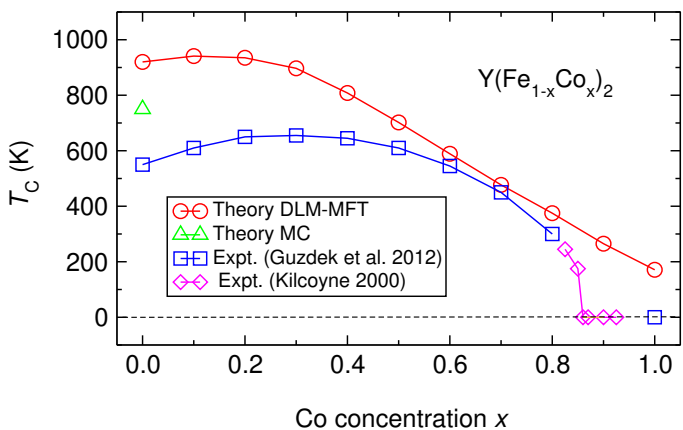}
\caption{\label{fig:DLM}
Curie temperatures of \yfeco{} as calculated with the DLM-MFT and MC methods compared with experimental data from works of Guzdek \textit{et al.}~\citep{guzdek_electrical_2012} and Kilcoyne~\citep{kilcoyne_evolution_2000}.
}
\end{figure}

\begin{table}[ht]
\centering
\caption{\label{tab:Curie} 
Curie temperatures (in K) of the \yfe, \yco, \zrfe{}, and \zrco{} as calculated
with the DLM-MFT and MC methods in comparison with experimental results from literature\citep{hilscher_low_1978,guzdek_electrical_2012}.
}
\vspace{1mm}
\begin{tabular}{|l|cccc|}
\hline \hline
	& \yfe{} &\yco{} & \zrfe{} &\zrco{} \\
\hline
DLM-MFT & 920 	& 171 	& 846   & 0 \\	
MC  	& 750	& -- 	& 780   & -- \\
Expt. 	& 550	& 0 	& 620   & 0 \\
\hline \hline
\end{tabular}
\end{table}
In Fig.~\ref{fig:DLM} we present our computational results for the Curie temperatures of the \yfeco{} system compared with the experiment. 
Table \ref{tab:Curie} summarizes the results for the boundary binary compounds.
For the \yfeco{} our calculations show qualitative agreement with the experiment for the concentration range $x\leq 0.8$ and preserve the characteristic Slater-Pauling-like maximum. 
A similar maximum also appears in the total magnetic moment vs Co concentration graph (not shown). 
The magnetic moments on Fe obtained with SPR-KKR, which are in good agreement with the experiment, are listed in Tab.~\ref{tab:mom}.
%
%
The Curie temperatures on the Fe-rich side calculated with the DLM-MFT are however significantly overestimated relative to the experimental values.
Similar disagreement between the critical temperatures measured and calculated with the mean field approximation has been found for other systems before and it was considered as a characteristic behavior of the mean field method~\citep{ebert_calculating_2011,ke_effects_2013,hedlund_magnetic_2017}.
The overestimation of $T_{\mathrm{C}}$ originates from the fact that the mean field approximation neglects the spin fluctuations which makes the magnetic moments more rigid~\cite{rusz_exchange_2006}.
%
%
On the Co-rich side, the predicted ferromagnetic ground state of \yco{} is wrong.
The comparison of non-magnetic and ferromagnetic states of \yco{} calculated in LDA at experimental lattice constants leads to an incorrect magnetic ground state.~\cite{khmelevskyi_magnetism_2005}
Furthermore, we observe that the formation of DLM state of \yco{} fails as the initially opposite magnetic moments on Co collapse to zero (non-magnetic state) after the convergence.
The problem with an accurate description of the magnetic state of Co alloys has been addressed previously by Edstr\"{o}em \textit{et al.}~\cite{edstrom_magnetic_2015}.
They have related the difficulties to an insufficient treatment of correlation effects in LDA/GGA, which can be overcome by application of the dynamical mean field theory (DMFT)~\cite{kotliar_electronic_2006}.
For YCo$_2$ we have made additional LDA+U calculations, with the corrections applied to Co 3$d$ orbitals.
However we have finished once again with a collapse of Co moments in the DLM state.
A failure of the LDA+U approach suggests the need for dynamical correlations in this case.
Similar collapse of magnetic moments on Co, as observed for \yco{}, we found for all DLM states of the considered \yfeco{} phases.
The overestimation of $T_{\mathrm{C}}$ in MFT on one hand and the insufficient treatment of correlation effects in LDA on the other are the reasons why the calculated $T_{\mathrm{C}}$ vs Co concentration dependence does not exhibit a ferromagnet-paramagnet phase transition.
\begin{figure}[ht]
\includegraphics[trim = 0 65 60 75,clip,height=\columnwidth, angle=270]{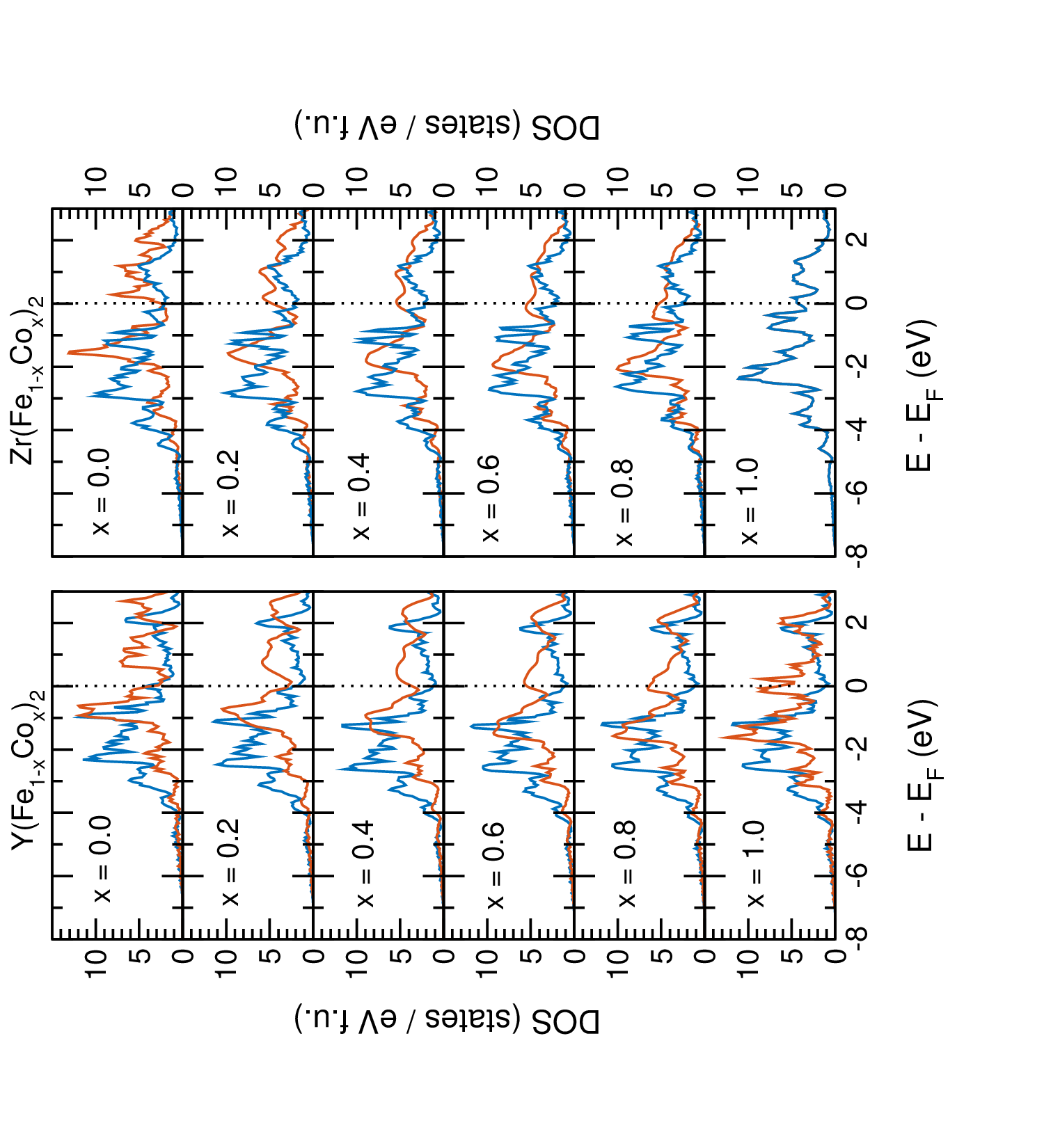}
\caption{\label{fig:DOS}
The densities of states (DOS) of \yfeco{} and \zrfeco{} as calculated with the FPLO5-LDA-CPA method. Red and blue colors represent two spin channels.
}
\end{figure}
The evolution of Curie temperature with Co concentration is correlated with evolution of an electronic structure of the material.
To give a feeling of what is happening as more Co atoms replace Fe ones in Fig.~\ref{fig:DOS} we present the total densities of states (DOS) for several Co concentrations.
The YFe$_2$ ($x = 0$) valence band consists mainly of Fe 3$d$ states.
The DOS of YFe$_2$ is strongly spin polarized and the most important contributions lay above -4~eV.
When alloying Co ($Z=27$) for Fe ($Z=26$) to the system are delivered the additional electrons, filling the valence band.
With the increase of Co concentration the less occupied spin channel is filling leading to decrease of spin polarization.
\begin{table}[ht]
\centering
\caption{\label{tab:mom} 
Magnetic moments  (in $\mu_B$) for \zrfe{} and \yfe{}, as used in UppASD MC simulations compared with the experimental results from literature\citep{yamada_nmr_1980}. Magnetic moments \textit{per Fe} include opposite contributions from 4$d$ elements and are equal to half of total magnetic moment per formula unit, 
while the calculated magnetic moments \textit{on Fe}
should be added together with
the opposite 4$d$ shares
to get the total magnetic moment.
}
\vspace{1mm}
\begin{tabular}{|c|cc|}
\hline \hline
  	& \yfe{} & \zrfe{} \\
\hline
on Fe (theory)  & 1.85 & 1.78 \\
per Fe (theory) & 1.66 & 1.58 \\	
per Fe (expt.) & 1.49 & 1.51  \\
\hline \hline
\end{tabular}
\end{table}

The magnetic moments on the Y/Zr atoms are of induced character and are expected to vanish with temperature, therefore we left them out of the simulated Heisenberg model.
\begin{figure}[ht]
\includegraphics[trim = 5 -15 0 40,clip,height=0.47\columnwidth, angle=0]{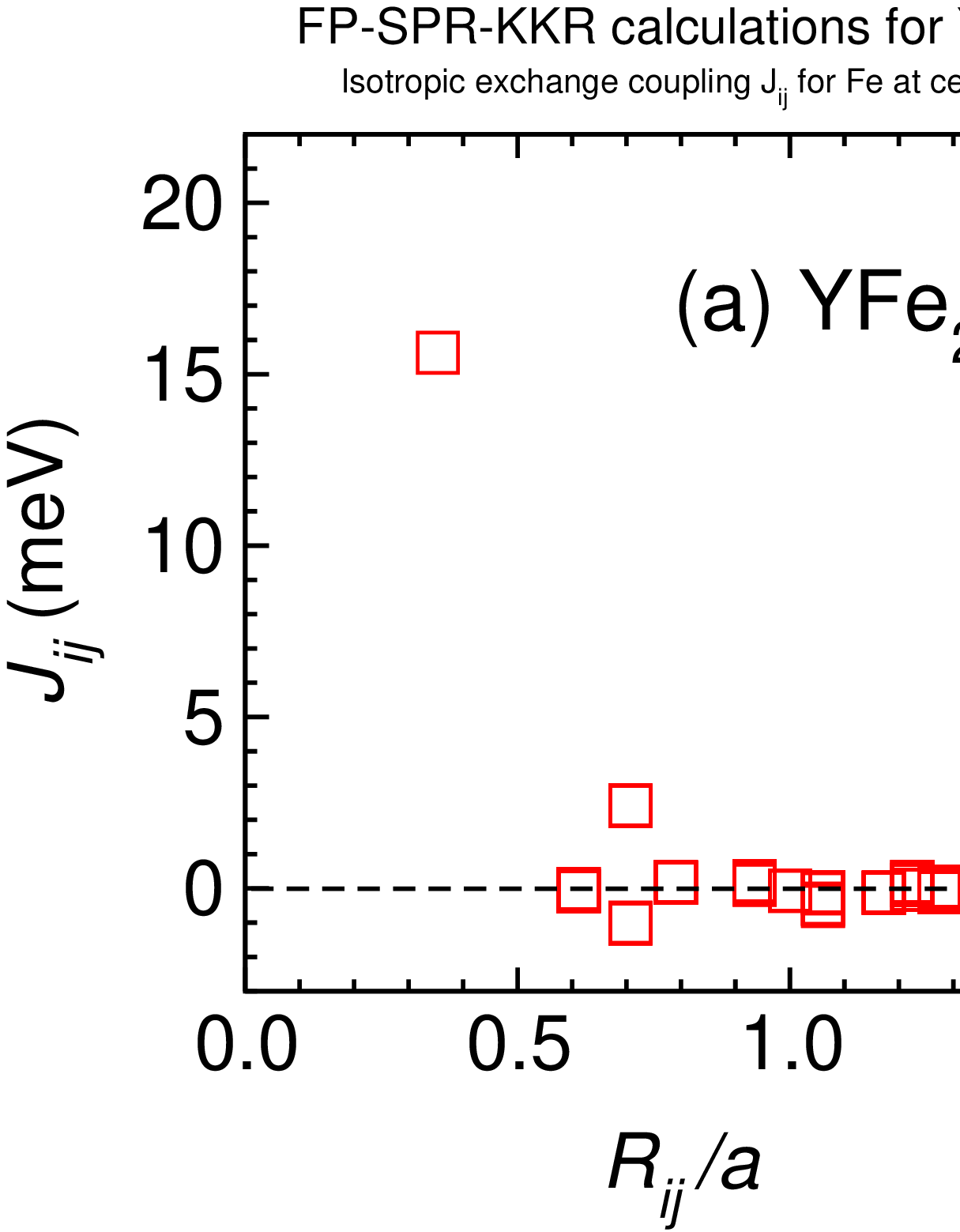}
\hfill
\includegraphics[trim = 5 5 0 30,clip,height=0.47\columnwidth, angle=0]{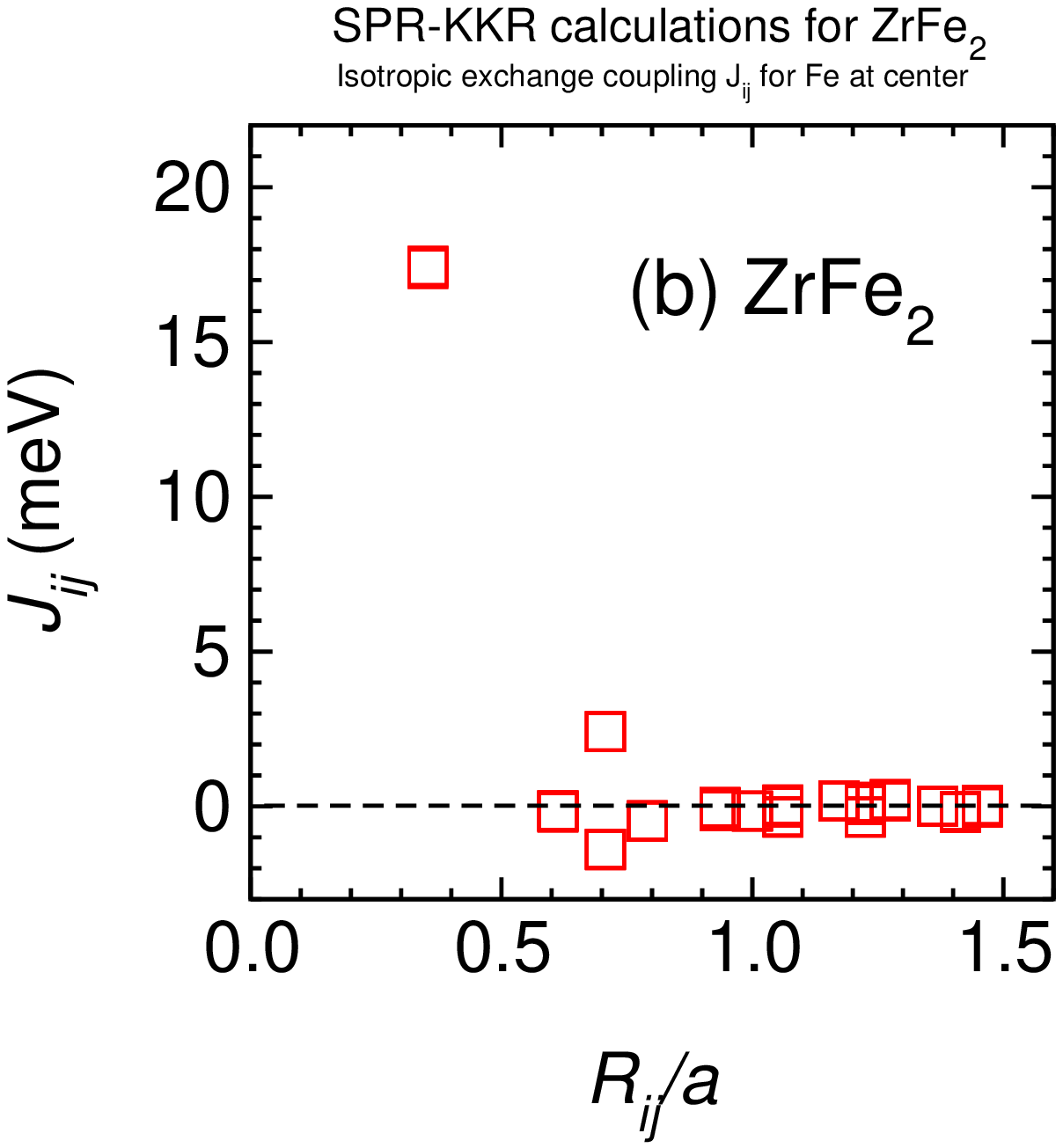}
\caption{\label{fig:YFeJXC}
The Fe-Fe exchange couplings vs normalized distance for (a) \yfe{} and (b) \zrfe{} as calculated with the FP-SPR-KKR.
}
\end{figure}
\begin{figure}[ht]
\includegraphics[trim = 80 5 0 40,clip,height=\columnwidth, angle=270]{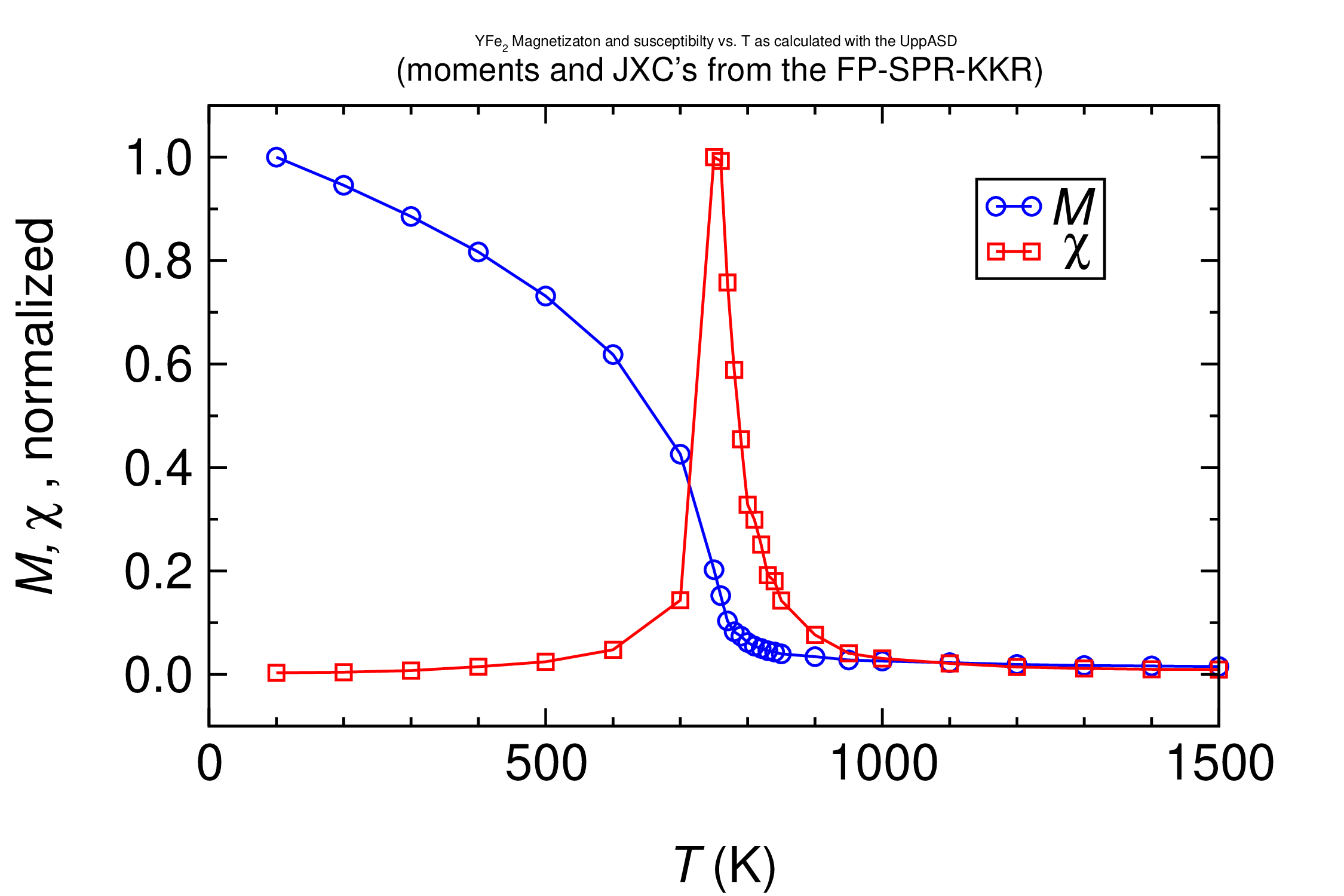}
\caption{\label{fig:YFe}
The normalized magnetization and susceptibility temperature dependencies for \yfe{} as calculated by MC simulations with the UppASD code with parameters from the FP-SPR-KKR.
}
\end{figure}
By looking at the graph of exchange interactions for \yfe{}, see Fig.~\ref{fig:YFeJXC}{(a)}, we can see that the dominant exchange coupling is positive, as expected for a ferromagnet. 
Besides direct exchange there can be also observed interactions of oscillatory character.
As of the MC simulations for the \yfe{}, see Fig.~\ref{fig:YFe}, the agreement with the experiment is reasonable where experimental \textit{T$_\mathrm{C}$} = $550 $~K and our simulations yield a value of \textit{T$_\mathrm{C}$}~=~750~K. 
We can also see that the $M(T)$ curve has a Curie-Weiss character, as expected.  

\subsection{Z\lowercase{r}(F\lowercase{e}$_{1-x}$C\lowercase{o}$_x$)$_{2}$}
\begin{figure}[ht]

\includegraphics[trim = 0 0 0 0,clip,width=\columnwidth]{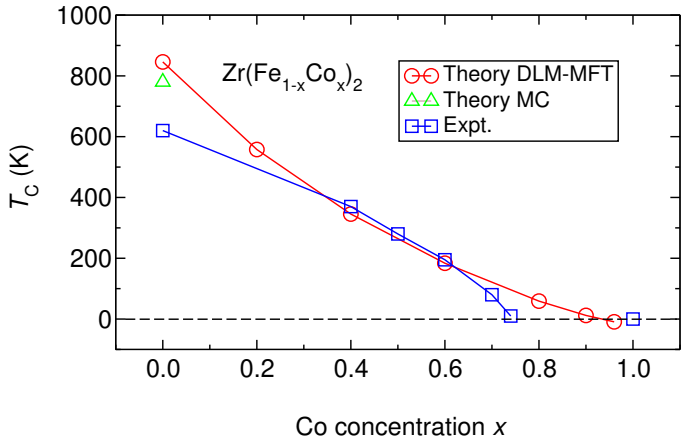}
\caption{\label{fig:DLM2}
Curie temperatures of \zrfeco{} as calculated with the DLM-MFT and MC methods in comparison with experimental data from literature\citep{hilscher_low_1978}.
}
\end{figure}

For the \zrfeco{} system, there is qualitative agreement of the calculated Curie temperatures with the experimental dependence, see Fig.~\ref{fig:DLM2}. The experiment has an almost linear course with $x$, ranging from $T_{\mathrm{C}}=620$~K\citep{hilscher_low_1978} at $x =0$ to 0~K at $x\sim 0.75$,
while our theoretical results also show that $T_{\mathrm{C}}$ is decreasing with $x$ and equal zero for $x\sim~ 0.95$. The discrepancy with the experiment in the Co-rich region can be explained similarly as before for \yfeco{}.
Furthermore, Fig.~\ref{fig:DOS} presents DOSs for several Co concentrations.
As Zr has one more electron than Y, the presented valence bands of \zrfeco{} alloys are slightly more filled than of corresponding \yfeco{} alloys.
The observed, for \zrfeco{} system, decrease of $T_C$ with $x$ is correlated with a decrease of spin splitting in DOS induced by band filling.
In agreement with experiment, we do not observe ferromagnetic ground state for ZrCo$_2$.

\begin{figure}[h!]
\includegraphics[trim = 100 5 0 90,clip,height=\columnwidth, angle=270]{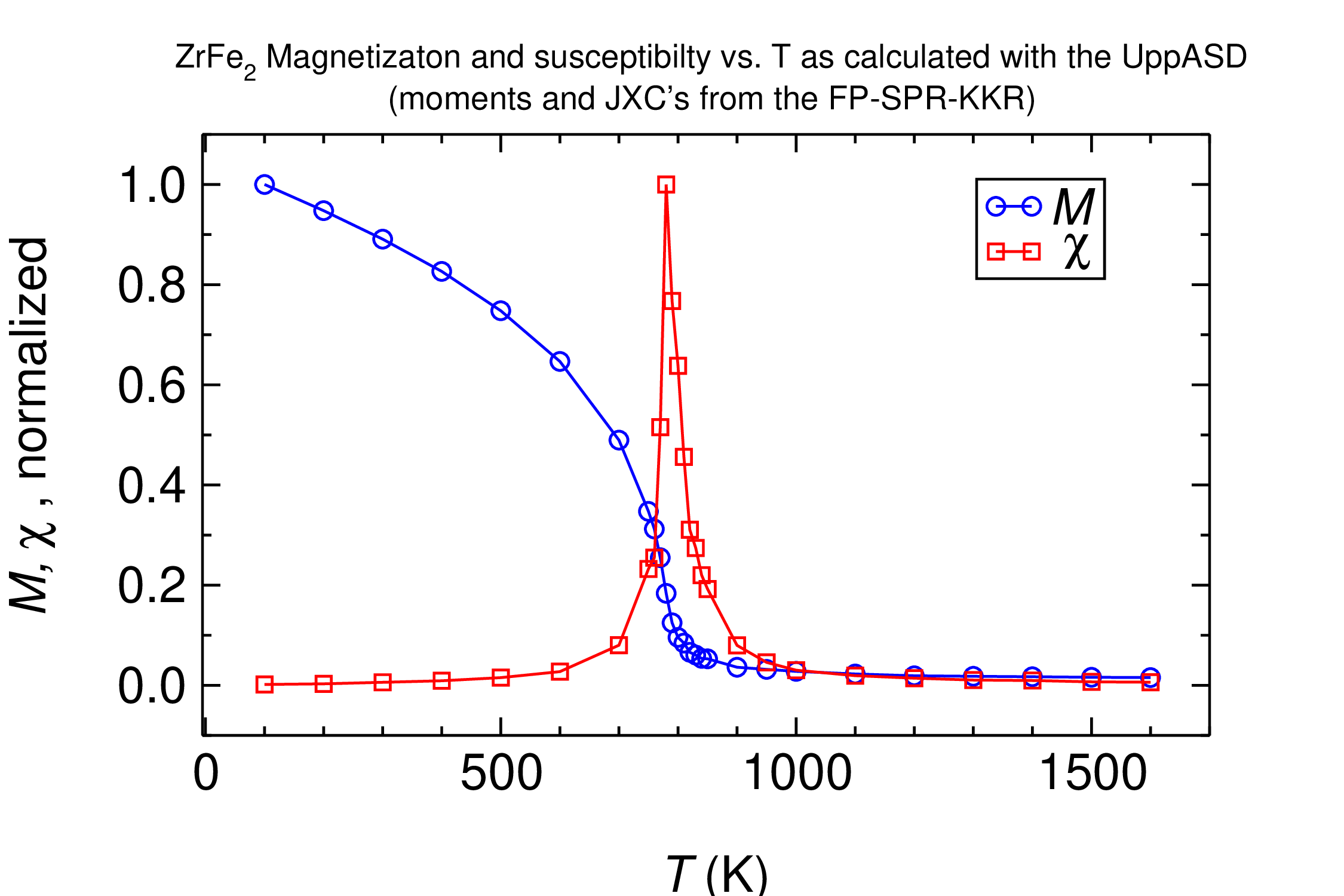}
\caption{\label{fig:ZrFe}
The normalized magnetization and susceptibility temperature dependencies  for \zrfe{} as calculated by MC simulations with the UppASD code with parameters from the FP-SPR-KKR.
}
\end{figure}
For the \zrfe{} the nearest neighbor Fe-Fe exchange coupling is positive and dominant. Similarly to \yfe{}, \zrfe{} exchange couplings also have oscillatory character, see Fig.~\ref{fig:YFeJXC}{(b)}. The \zrfe{} magnetization and susceptibility MC simulations, see Fig.~\ref{fig:ZrFe}, predict a ferromagnetic state for \zrfe{} with the Curie temperature of ~$\sim 780$~K which is in a reasonable agreement with the experimental value of ~$620$~K. Interestingly for the \zrfe{} the MC simulations with the dominant exchange coupling only yielded $T_\mathrm{C}~\sim 550$~K (230~K lower than for exchange couplings considered up to 1.5~\textit{a}) which shows how the system is sensitive to small contributions in relatively large numbers and the importance of checking for convergence with the radius of the exchange integrals cutoff sphere. Analogously to \yfe{}, the \zrfe{} $M(T)$ curve also has a Curie-Weiss character.
We have decided not to model \yco{} and \zrco{} with MC simulations due to the fact
that the exchange couplings and magnetic moments used in the mentioned calculations were obtained with respect to the ferromagnetic ground state and \yco{} and \zrco{} are Pauli paramagnets.

\section{Summary and Conclusions}

The Curie temperatures of \yfeco{} and \zrfeco{} systems were calculated \textit{ab initio} based on the disordered local moment method within the mean field theory (DLM-MFT).
Furthermore, the Curie temperatures of \yfe{} and \zrfe{} were calculated by Monte Carlo (MC) simulations.
Comparing the results of our calculations with experimental data from literature, a good qualitative agreement is observed.
The main features of experimental $T_{\mathrm{C}}$($x$) plots are reproduced, however with some limitations.
Both the DLM-MFT and MC tend to overestimate the $T_{\mathrm{C}}$ with the most troublesome region on the Co-rich side, where the systems undergo a ferromagnetic-paramagnetic phase transition. 
For \yfeco{} the DLM-MFT approach correctly predicts the characteristic maximum in $T_{\mathrm{C}}$ for intermediate compositions but fails by predicting the ferromagnetic ground state of the \yco{} paramagnet.
For \zrfeco{} the DLM-MFT predicts both the monotonic decrease of $T_{\mathrm{C}}$ with $x$ and the critical Co concentration above which the system becomes paramagnetic.
However, the exact value of the critical concentration is overestimated.
The MC simulations for \yfe{} and \zrfe{} indicate that the nearest neighbors Fe-Fe exchange interactions are the most responsible for the value of $T_{\mathrm{C}}$.
From the point of view of computations this paper shows that the accurate $T_{\mathrm{C}}$ analysis for alloys is feasible with coherent potential approximation used for simulation of the magnetically disordered DLM state.

\section{Acknowledgements}

We acknowledge the financial support from the Foundation of Polish 
Science grant HOMING.
The HOMING programme is co-financed by the European Union under the 
European Regional Development Fund.
Part of the computations were performed on resources provided by the Poznań Supercomputing and Networking Center (PSNC).
We also thank Dr. Z. \'Sniadecki for comments and discussion.

\end{sloppypar}

\bibliography{ZrYFeCo}    

\begin{thebibliography}{43}%
\makeatletter
\providecommand \@ifxundefined [1]{%
 \@ifx{#1\undefined}
}%
\providecommand \@ifnum [1]{%
 \ifnum #1\expandafter \@firstoftwo
 \else \expandafter \@secondoftwo
 \fi
}%
\providecommand \@ifx [1]{%
 \ifx #1\expandafter \@firstoftwo
 \else \expandafter \@secondoftwo
 \fi
}%
\providecommand \natexlab [1]{#1}%
\providecommand \enquote  [1]{``#1''}%
\providecommand \bibnamefont  [1]{#1}%
\providecommand \bibfnamefont [1]{#1}%
\providecommand \citenamefont [1]{#1}%
\providecommand \href@noop [0]{\@secondoftwo}%
\providecommand \href [0]{\begingroup \@sanitize@url \@href}%
\providecommand \@href[1]{\@@startlink{#1}\@@href}%
\providecommand \@@href[1]{\endgroup#1\@@endlink}%
\providecommand \@sanitize@url [0]{\catcode `\\12\catcode `\$12\catcode
  `\&12\catcode `\#12\catcode `\^12\catcode `\_12\catcode `\%12\relax}%
\providecommand \@@startlink[1]{}%
\providecommand \@@endlink[0]{}%
\providecommand \url  [0]{\begingroup\@sanitize@url \@url }%
\providecommand \@url [1]{\endgroup\@href {#1}{\urlprefix }}%
\providecommand \urlprefix  [0]{URL }%
\providecommand \Eprint [0]{\href }%
\providecommand \doibase [0]{http://dx.doi.org/}%
\providecommand \selectlanguage [0]{\@gobble}%
\providecommand \bibinfo  [0]{\@secondoftwo}%
\providecommand \bibfield  [0]{\@secondoftwo}%
\providecommand \translation [1]{[#1]}%
\providecommand \BibitemOpen [0]{}%
\providecommand \bibitemStop [0]{}%
\providecommand \bibitemNoStop [0]{.\EOS\space}%
\providecommand \EOS [0]{\spacefactor3000\relax}%
\providecommand \BibitemShut  [1]{\csname bibitem#1\endcsname}%
\let\auto@bib@innerbib\@empty
\bibitem [{\citenamefont {Wiesinger}\ \emph {et~al.}(2005)\citenamefont
  {Wiesinger}, \citenamefont {Paul-Boncour}, \citenamefont {Filipek},
  \citenamefont {Reichl}, \citenamefont {Marchuk},\ and\ \citenamefont
  {Percheron-Gu{\'e}gan}}]{wiesinger_structural_2005}%
  \BibitemOpen
  \bibfield  {author} {\bibinfo {author} {\bibfnamefont {G.}~\bibnamefont
  {Wiesinger}}, \bibinfo {author} {\bibfnamefont {V.}~\bibnamefont
  {Paul-Boncour}}, \bibinfo {author} {\bibfnamefont {S.~M.}\ \bibnamefont
  {Filipek}}, \bibinfo {author} {\bibfnamefont {C.}~\bibnamefont {Reichl}},
  \bibinfo {author} {\bibfnamefont {I.}~\bibnamefont {Marchuk}}, \ and\
  \bibinfo {author} {\bibfnamefont {A.}~\bibnamefont {Percheron-Gu{\'e}gan}},\
  }\href {\doibase 10.1088/0953-8984/17/6/009} {\bibfield  {journal} {\bibinfo
  {journal} {Journal of Physics: Condensed Matter}\ }\textbf {\bibinfo {volume}
  {17}},\ \bibinfo {pages} {893} (\bibinfo {year} {2005})}\BibitemShut
  {NoStop}%
\bibitem [{\citenamefont {Isnard}\ \emph {et~al.}(2011)\citenamefont {Isnard},
  \citenamefont {Paul-Boncour}, \citenamefont {Arnold}, \citenamefont {Colin},
  \citenamefont {Leblond}, \citenamefont {Kamarad},\ and\ \citenamefont
  {Sugiura}}]{isnard_pressure-induced_2011}%
  \BibitemOpen
  \bibfield  {author} {\bibinfo {author} {\bibfnamefont {O.}~\bibnamefont
  {Isnard}}, \bibinfo {author} {\bibfnamefont {V.}~\bibnamefont
  {Paul-Boncour}}, \bibinfo {author} {\bibfnamefont {Z.}~\bibnamefont
  {Arnold}}, \bibinfo {author} {\bibfnamefont {C.~V.}\ \bibnamefont {Colin}},
  \bibinfo {author} {\bibfnamefont {T.}~\bibnamefont {Leblond}}, \bibinfo
  {author} {\bibfnamefont {J.}~\bibnamefont {Kamarad}}, \ and\ \bibinfo
  {author} {\bibfnamefont {H.}~\bibnamefont {Sugiura}},\ }\href {\doibase
  10.1103/PhysRevB.84.094429} {\bibfield  {journal} {\bibinfo  {journal}
  {Physical Review B}\ }\textbf {\bibinfo {volume} {84}},\ \bibinfo {pages}
  {094429} (\bibinfo {year} {2011})}\BibitemShut {NoStop}%
\bibitem [{\citenamefont {Li}\ \emph {et~al.}(2016)\citenamefont {Li},
  \citenamefont {Wang}, \citenamefont {Ouyang}, \citenamefont {Liu},\ and\
  \citenamefont {Zhu}}]{li_reversible_2016}%
  \BibitemOpen
  \bibfield  {author} {\bibinfo {author} {\bibfnamefont {Z.}~\bibnamefont
  {Li}}, \bibinfo {author} {\bibfnamefont {H.}~\bibnamefont {Wang}}, \bibinfo
  {author} {\bibfnamefont {L.}~\bibnamefont {Ouyang}}, \bibinfo {author}
  {\bibfnamefont {J.}~\bibnamefont {Liu}}, \ and\ \bibinfo {author}
  {\bibfnamefont {M.}~\bibnamefont {Zhu}},\ }\href {\doibase
  10.1016/j.jallcom.2016.08.003} {\bibfield  {journal} {\bibinfo  {journal}
  {Journal of Alloys and Compounds}\ }\textbf {\bibinfo {volume} {689}},\
  \bibinfo {pages} {843} (\bibinfo {year} {2016})}\BibitemShut {NoStop}%
\bibitem [{\citenamefont {Jat}\ \emph {et~al.}(2015)\citenamefont {Jat},
  \citenamefont {Singh}, \citenamefont {Parida}, \citenamefont {Das},
  \citenamefont {Agarwal}, \citenamefont {Mukerjee},\ and\ \citenamefont
  {Ramakumar}}]{jat_structural_2015}%
  \BibitemOpen
  \bibfield  {author} {\bibinfo {author} {\bibfnamefont {R.~A.}\ \bibnamefont
  {Jat}}, \bibinfo {author} {\bibfnamefont {R.}~\bibnamefont {Singh}}, \bibinfo
  {author} {\bibfnamefont {S.~C.}\ \bibnamefont {Parida}}, \bibinfo {author}
  {\bibfnamefont {A.}~\bibnamefont {Das}}, \bibinfo {author} {\bibfnamefont
  {R.}~\bibnamefont {Agarwal}}, \bibinfo {author} {\bibfnamefont {S.~K.}\
  \bibnamefont {Mukerjee}}, \ and\ \bibinfo {author} {\bibfnamefont {K.~L.}\
  \bibnamefont {Ramakumar}},\ }\href {\doibase 10.1016/j.ijhydene.2015.02.094}
  {\bibfield  {journal} {\bibinfo  {journal} {International Journal of Hydrogen
  Energy}\ }\textbf {\bibinfo {volume} {40}},\ \bibinfo {pages} {5135}
  (\bibinfo {year} {2015})}\BibitemShut {NoStop}%
\bibitem [{\citenamefont {Merlino}\ \emph {et~al.}(2016)\citenamefont
  {Merlino}, \citenamefont {Luna}, \citenamefont {Juan},\ and\ \citenamefont
  {Pronsato}}]{merlino_dft_2016}%
  \BibitemOpen
  \bibfield  {author} {\bibinfo {author} {\bibfnamefont {A.~R.}\ \bibnamefont
  {Merlino}}, \bibinfo {author} {\bibfnamefont {C.~R.}\ \bibnamefont {Luna}},
  \bibinfo {author} {\bibfnamefont {A.}~\bibnamefont {Juan}}, \ and\ \bibinfo
  {author} {\bibfnamefont {M.~E.}\ \bibnamefont {Pronsato}},\ }\href {\doibase
  10.1016/j.ijhydene.2015.10.077} {\bibfield  {journal} {\bibinfo  {journal}
  {International Journal of Hydrogen Energy}\ }\textbf {\bibinfo {volume}
  {41}},\ \bibinfo {pages} {2700} (\bibinfo {year} {2016})}\BibitemShut
  {NoStop}%
\bibitem [{\citenamefont {Shahi}\ \emph {et~al.}(2015)\citenamefont {Shahi},
  \citenamefont {Bhatanagar}, \citenamefont {Pandey}, \citenamefont {Shukla},
  \citenamefont {Yadav}, \citenamefont {Shaz},\ and\ \citenamefont
  {Srivastava}}]{shahi_mgh2zrfe2hx_2015}%
  \BibitemOpen
  \bibfield  {author} {\bibinfo {author} {\bibfnamefont {R.~R.}\ \bibnamefont
  {Shahi}}, \bibinfo {author} {\bibfnamefont {A.}~\bibnamefont {Bhatanagar}},
  \bibinfo {author} {\bibfnamefont {S.~K.}\ \bibnamefont {Pandey}}, \bibinfo
  {author} {\bibfnamefont {V.}~\bibnamefont {Shukla}}, \bibinfo {author}
  {\bibfnamefont {T.~P.}\ \bibnamefont {Yadav}}, \bibinfo {author}
  {\bibfnamefont {M.~A.}\ \bibnamefont {Shaz}}, \ and\ \bibinfo {author}
  {\bibfnamefont {O.~N.}\ \bibnamefont {Srivastava}},\ }\href {\doibase
  10.1016/j.ijhydene.2015.03.162} {\bibfield  {journal} {\bibinfo  {journal}
  {International Journal of Hydrogen Energy}\ }\textbf {\bibinfo {volume}
  {40}},\ \bibinfo {pages} {11506} (\bibinfo {year} {2015})}\BibitemShut
  {NoStop}%
\bibitem [{\citenamefont {Shukla}\ \emph {et~al.}(2017)\citenamefont {Shukla},
  \citenamefont {Bhatnagar}, \citenamefont {K.~Soni}, \citenamefont
  {K.~Vishwakarma}, \citenamefont {A.~Shaz}, \citenamefont {P.~Yadav},\ and\
  \citenamefont {N.~Srivastava}}]{shukla_enhanced_2017}%
  \BibitemOpen
  \bibfield  {author} {\bibinfo {author} {\bibfnamefont {V.}~\bibnamefont
  {Shukla}}, \bibinfo {author} {\bibfnamefont {A.}~\bibnamefont {Bhatnagar}},
  \bibinfo {author} {\bibfnamefont {P.}~\bibnamefont {K.~Soni}}, \bibinfo
  {author} {\bibfnamefont {A.}~\bibnamefont {K.~Vishwakarma}}, \bibinfo
  {author} {\bibfnamefont {M.}~\bibnamefont {A.~Shaz}}, \bibinfo {author}
  {\bibfnamefont {T.}~\bibnamefont {P.~Yadav}}, \ and\ \bibinfo {author}
  {\bibfnamefont {O.}~\bibnamefont {N.~Srivastava}},\ }\href {\doibase
  10.1039/C6CP08333A} {\bibfield  {journal} {\bibinfo  {journal} {Physical
  Chemistry Chemical Physics}\ }\textbf {\bibinfo {volume} {19}},\ \bibinfo
  {pages} {9444} (\bibinfo {year} {2017})}\BibitemShut {NoStop}%
\bibitem [{\citenamefont {Baranov}\ \emph {et~al.}(2009)\citenamefont
  {Baranov}, \citenamefont {Proshkin}, \citenamefont {Czternasty},
  \citenamefont {Mei{\ss}ner}, \citenamefont {Podlesnyak},\ and\ \citenamefont
  {Podgornykh}}]{baranov_butterflylike_2009}%
  \BibitemOpen
  \bibfield  {author} {\bibinfo {author} {\bibfnamefont {N.~V.}\ \bibnamefont
  {Baranov}}, \bibinfo {author} {\bibfnamefont {A.~V.}\ \bibnamefont
  {Proshkin}}, \bibinfo {author} {\bibfnamefont {C.}~\bibnamefont
  {Czternasty}}, \bibinfo {author} {\bibfnamefont {M.}~\bibnamefont
  {Mei{\ss}ner}}, \bibinfo {author} {\bibfnamefont {A.}~\bibnamefont
  {Podlesnyak}}, \ and\ \bibinfo {author} {\bibfnamefont {S.~M.}\ \bibnamefont
  {Podgornykh}},\ }\href {\doibase 10.1103/PhysRevB.79.184420} {\bibfield
  {journal} {\bibinfo  {journal} {Physical Review B}\ }\textbf {\bibinfo
  {volume} {79}} (\bibinfo {year} {2009}),\
  10.1103/PhysRevB.79.184420}\BibitemShut {NoStop}%
\bibitem [{\citenamefont {Pierunek}\ \emph {et~al.}(2017)\citenamefont
  {Pierunek}, \citenamefont {{\'S}niadecki}, \citenamefont {Werwi{\'n}ski},
  \citenamefont {Wasilewski}, \citenamefont {Franco},\ and\ \citenamefont
  {Idzikowski}}]{pierunek_normal_2017}%
  \BibitemOpen
  \bibfield  {author} {\bibinfo {author} {\bibfnamefont {N.}~\bibnamefont
  {Pierunek}}, \bibinfo {author} {\bibfnamefont {Z.}~\bibnamefont
  {{\'S}niadecki}}, \bibinfo {author} {\bibfnamefont {M.}~\bibnamefont
  {Werwi{\'n}ski}}, \bibinfo {author} {\bibfnamefont {B.}~\bibnamefont
  {Wasilewski}}, \bibinfo {author} {\bibfnamefont {V.}~\bibnamefont {Franco}},
  \ and\ \bibinfo {author} {\bibfnamefont {B.}~\bibnamefont {Idzikowski}},\
  }\href {\doibase 10.1016/j.jallcom.2017.01.181} {\bibfield  {journal}
  {\bibinfo  {journal} {Journal of Alloys and Compounds}\ }\textbf {\bibinfo
  {volume} {702}},\ \bibinfo {pages} {258} (\bibinfo {year}
  {2017})}\BibitemShut {NoStop}%
\bibitem [{\citenamefont {Mican}\ \emph {et~al.}(2013)\citenamefont {Mican},
  \citenamefont {Benea}, \citenamefont {Mankovsky}, \citenamefont {Polesya},
  \citenamefont {G{\^i}nsc{\u a}},\ and\ \citenamefont
  {Tetean}}]{mican_magnetic_2013}%
  \BibitemOpen
  \bibfield  {author} {\bibinfo {author} {\bibfnamefont {S.}~\bibnamefont
  {Mican}}, \bibinfo {author} {\bibfnamefont {D.}~\bibnamefont {Benea}},
  \bibinfo {author} {\bibfnamefont {S.}~\bibnamefont {Mankovsky}}, \bibinfo
  {author} {\bibfnamefont {S.}~\bibnamefont {Polesya}}, \bibinfo {author}
  {\bibfnamefont {O.}~\bibnamefont {G{\^i}nsc{\u a}}}, \ and\ \bibinfo {author}
  {\bibfnamefont {R.}~\bibnamefont {Tetean}},\ }\href {\doibase
  10.1088/0953-8984/25/46/466003} {\bibfield  {journal} {\bibinfo  {journal}
  {Journal of Physics: Condensed Matter}\ }\textbf {\bibinfo {volume} {25}},\
  \bibinfo {pages} {466003} (\bibinfo {year} {2013})}\BibitemShut {NoStop}%
\bibitem [{\citenamefont {Chen}\ \emph {et~al.}(2015)\citenamefont {Chen},
  \citenamefont {Sun}, \citenamefont {Duan}, \citenamefont {Huang},\ and\
  \citenamefont {Peng}}]{chen_phase_2015}%
  \BibitemOpen
  \bibfield  {author} {\bibinfo {author} {\bibfnamefont {S.}~\bibnamefont
  {Chen}}, \bibinfo {author} {\bibfnamefont {Y.}~\bibnamefont {Sun}}, \bibinfo
  {author} {\bibfnamefont {Y.-H.}\ \bibnamefont {Duan}}, \bibinfo {author}
  {\bibfnamefont {B.}~\bibnamefont {Huang}}, \ and\ \bibinfo {author}
  {\bibfnamefont {M.-J.}\ \bibnamefont {Peng}},\ }\href {\doibase
  10.1016/j.jallcom.2015.01.038} {\bibfield  {journal} {\bibinfo  {journal}
  {Journal of Alloys and Compounds}\ }\textbf {\bibinfo {volume} {630}},\
  \bibinfo {pages} {202} (\bibinfo {year} {2015})}\BibitemShut {NoStop}%
\bibitem [{\citenamefont {Bhatt}\ \emph {et~al.}(2016)\citenamefont {Bhatt},
  \citenamefont {Kumar}, \citenamefont {Arora}, \citenamefont {Bapna},\ and\
  \citenamefont {Ahuja}}]{bhatt_high_2016}%
  \BibitemOpen
  \bibfield  {author} {\bibinfo {author} {\bibfnamefont {S.}~\bibnamefont
  {Bhatt}}, \bibinfo {author} {\bibfnamefont {K.}~\bibnamefont {Kumar}},
  \bibinfo {author} {\bibfnamefont {G.}~\bibnamefont {Arora}}, \bibinfo
  {author} {\bibfnamefont {K.}~\bibnamefont {Bapna}}, \ and\ \bibinfo {author}
  {\bibfnamefont {B.~L.}\ \bibnamefont {Ahuja}},\ }\href {\doibase
  10.1016/j.radphyschem.2016.03.021} {\bibfield  {journal} {\bibinfo  {journal}
  {Radiation Physics and Chemistry}\ }\textbf {\bibinfo {volume} {125}},\
  \bibinfo {pages} {109} (\bibinfo {year} {2016})}\BibitemShut {NoStop}%
\bibitem [{\citenamefont {Zhang}\ and\ \citenamefont
  {Zhang}(2015)}]{zhang_prediction_2015}%
  \BibitemOpen
  \bibfield  {author} {\bibinfo {author} {\bibfnamefont {W.}~\bibnamefont
  {Zhang}}\ and\ \bibinfo {author} {\bibfnamefont {W.}~\bibnamefont {Zhang}},\
  }\href {\doibase 10.1063/1.4919424} {\bibfield  {journal} {\bibinfo
  {journal} {Journal of Applied Physics}\ }\textbf {\bibinfo {volume} {117}},\
  \bibinfo {pages} {163917} (\bibinfo {year} {2015})}\BibitemShut {NoStop}%
\bibitem [{\citenamefont {Zhuravleva}\ \emph {et~al.}(2017)\citenamefont
  {Zhuravleva}, \citenamefont {Grechnev}, \citenamefont {Panfilov},\ and\
  \citenamefont {Lyogenkaya}}]{zhuravleva_magnetovolume_2017}%
  \BibitemOpen
  \bibfield  {author} {\bibinfo {author} {\bibfnamefont {I.~P.}\ \bibnamefont
  {Zhuravleva}}, \bibinfo {author} {\bibfnamefont {G.~E.}\ \bibnamefont
  {Grechnev}}, \bibinfo {author} {\bibfnamefont {A.~S.}\ \bibnamefont
  {Panfilov}}, \ and\ \bibinfo {author} {\bibfnamefont {A.~A.}\ \bibnamefont
  {Lyogenkaya}},\ }\href {\doibase 10.1063/1.4985217} {\bibfield  {journal}
  {\bibinfo  {journal} {Low Temperature Physics}\ }\textbf {\bibinfo {volume}
  {43}},\ \bibinfo {pages} {597} (\bibinfo {year} {2017})}\BibitemShut
  {NoStop}%
\bibitem [{\citenamefont {Kumar}\ \emph {et~al.}(2014)\citenamefont {Kumar},
  \citenamefont {Kashyap}, \citenamefont {Balamurugan}, \citenamefont {Shield},
  \citenamefont {Sellmyer},\ and\ \citenamefont
  {Skomski}}]{kumar_permanent_2014}%
  \BibitemOpen
  \bibfield  {author} {\bibinfo {author} {\bibfnamefont {P.}~\bibnamefont
  {Kumar}}, \bibinfo {author} {\bibfnamefont {A.}~\bibnamefont {Kashyap}},
  \bibinfo {author} {\bibfnamefont {B.}~\bibnamefont {Balamurugan}}, \bibinfo
  {author} {\bibfnamefont {J.~E.}\ \bibnamefont {Shield}}, \bibinfo {author}
  {\bibfnamefont {D.~J.}\ \bibnamefont {Sellmyer}}, \ and\ \bibinfo {author}
  {\bibfnamefont {R.}~\bibnamefont {Skomski}},\ }\href {\doibase
  10.1088/0953-8984/26/6/064209} {\bibfield  {journal} {\bibinfo  {journal}
  {Journal of Physics: Condensed Matter}\ }\textbf {\bibinfo {volume} {26}},\
  \bibinfo {pages} {064209} (\bibinfo {year} {2014})}\BibitemShut {NoStop}%
\bibitem [{\citenamefont {Hilscher}\ and\ \citenamefont
  {Gmelin}(1978)}]{hilscher_low_1978}%
  \BibitemOpen
  \bibfield  {author} {\bibinfo {author} {\bibfnamefont {G.}~\bibnamefont
  {Hilscher}}\ and\ \bibinfo {author} {\bibfnamefont {E.}~\bibnamefont
  {Gmelin}},\ }\href
  {http://jphyscol.journaldephysique.org/articles/jphyscol/abs/1978/06/jphyscol197839C6345/jphyscol197839C6345.html}
  {\bibfield  {journal} {\bibinfo  {journal} {Le Journal de Physique
  Colloques}\ }\textbf {\bibinfo {volume} {39}},\ \bibinfo {pages} {C6}
  (\bibinfo {year} {1978})}\BibitemShut {NoStop}%
\bibitem [{\citenamefont {Guzdek}\ \emph {et~al.}(2012)\citenamefont {Guzdek},
  \citenamefont {Pszczo{\l }a}, \citenamefont {Chmist}, \citenamefont {Stoch},
  \citenamefont {Stoch},\ and\ \citenamefont
  {Suwalski}}]{guzdek_electrical_2012}%
  \BibitemOpen
  \bibfield  {author} {\bibinfo {author} {\bibfnamefont {P.}~\bibnamefont
  {Guzdek}}, \bibinfo {author} {\bibfnamefont {J.}~\bibnamefont {Pszczo{\l
  }a}}, \bibinfo {author} {\bibfnamefont {J.}~\bibnamefont {Chmist}}, \bibinfo
  {author} {\bibfnamefont {P.}~\bibnamefont {Stoch}}, \bibinfo {author}
  {\bibfnamefont {A.}~\bibnamefont {Stoch}}, \ and\ \bibinfo {author}
  {\bibfnamefont {J.}~\bibnamefont {Suwalski}},\ }\href {\doibase
  10.1016/j.jallcom.2011.12.081} {\bibfield  {journal} {\bibinfo  {journal}
  {Journal of Alloys and Compounds}\ }\textbf {\bibinfo {volume} {520}},\
  \bibinfo {pages} {72} (\bibinfo {year} {2012})}\BibitemShut {NoStop}%
\bibitem [{\citenamefont {Yamada}\ and\ \citenamefont
  {Ohmae}(1980)}]{yamada_nmr_1980}%
  \BibitemOpen
  \bibfield  {author} {\bibinfo {author} {\bibfnamefont {Y.}~\bibnamefont
  {Yamada}}\ and\ \bibinfo {author} {\bibfnamefont {H.}~\bibnamefont {Ohmae}},\
  }\href {\doibase 10.1143/JPSJ.48.1513} {\bibfield  {journal} {\bibinfo
  {journal} {Journal of the Physical Society of Japan}\ }\textbf {\bibinfo
  {volume} {48}},\ \bibinfo {pages} {1513} (\bibinfo {year}
  {1980})}\BibitemShut {NoStop}%
\bibitem [{\citenamefont {Goto}\ \emph {et~al.}(1990)\citenamefont {Goto},
  \citenamefont {Sakakibara}, \citenamefont {Murata}, \citenamefont {Komatsu},\
  and\ \citenamefont {Fukamichi}}]{goto_itinerant_1990}%
  \BibitemOpen
  \bibfield  {author} {\bibinfo {author} {\bibfnamefont {T.}~\bibnamefont
  {Goto}}, \bibinfo {author} {\bibfnamefont {T.}~\bibnamefont {Sakakibara}},
  \bibinfo {author} {\bibfnamefont {K.}~\bibnamefont {Murata}}, \bibinfo
  {author} {\bibfnamefont {H.}~\bibnamefont {Komatsu}}, \ and\ \bibinfo
  {author} {\bibfnamefont {K.}~\bibnamefont {Fukamichi}},\ }\href {\doibase
  10.1016/S0304-8853(10)80256-2} {\bibfield  {journal} {\bibinfo  {journal}
  {Journal of Magnetism and Magnetic Materials}\ }\textbf {\bibinfo {volume}
  {90}},\ \bibinfo {pages} {700} (\bibinfo {year} {1990})}\BibitemShut
  {NoStop}%
\bibitem [{\citenamefont {Yamada}\ and\ \citenamefont
  {Shimizu}(1990)}]{yamada_itinerant_1990}%
  \BibitemOpen
  \bibfield  {author} {\bibinfo {author} {\bibfnamefont {H.}~\bibnamefont
  {Yamada}}\ and\ \bibinfo {author} {\bibfnamefont {M.}~\bibnamefont
  {Shimizu}},\ }\href
  {http://www.sciencedirect.com/science/article/pii/S0304885310802574}
  {\bibfield  {journal} {\bibinfo  {journal} {Journal of Magnetism and Magnetic
  Materials}\ }\textbf {\bibinfo {volume} {90}},\ \bibinfo {pages} {703}
  (\bibinfo {year} {1990})}\BibitemShut {NoStop}%
\bibitem [{\citenamefont {{\'S}niadecki}\ \emph {et~al.}(2014)\citenamefont
  {{\'S}niadecki}, \citenamefont {Werwi{\'n}ski}, \citenamefont {Szajek},
  \citenamefont {R{\"o}{\ss}ler},\ and\ \citenamefont
  {Idzikowski}}]{sniadecki_induced_2014}%
  \BibitemOpen
  \bibfield  {author} {\bibinfo {author} {\bibfnamefont {Z.}~\bibnamefont
  {{\'S}niadecki}}, \bibinfo {author} {\bibfnamefont {M.}~\bibnamefont
  {Werwi{\'n}ski}}, \bibinfo {author} {\bibfnamefont {A.}~\bibnamefont
  {Szajek}}, \bibinfo {author} {\bibfnamefont {U.~K.}\ \bibnamefont
  {R{\"o}{\ss}ler}}, \ and\ \bibinfo {author} {\bibfnamefont {B.}~\bibnamefont
  {Idzikowski}},\ }\href {\doibase 10.1063/1.4866848} {\bibfield  {journal}
  {\bibinfo  {journal} {Journal of Applied Physics}\ }\textbf {\bibinfo
  {volume} {115}},\ \bibinfo {pages} {17E129} (\bibinfo {year}
  {2014})}\BibitemShut {NoStop}%
\bibitem [{\citenamefont {{\'S}niadecki}\ \emph {et~al.}(2015)\citenamefont
  {{\'S}niadecki}, \citenamefont {Kopcewicz}, \citenamefont {Pierunek},\ and\
  \citenamefont {Idzikowski}}]{sniadecki_magnetic_2015}%
  \BibitemOpen
  \bibfield  {author} {\bibinfo {author} {\bibfnamefont {Z.}~\bibnamefont
  {{\'S}niadecki}}, \bibinfo {author} {\bibfnamefont {M.}~\bibnamefont
  {Kopcewicz}}, \bibinfo {author} {\bibfnamefont {N.}~\bibnamefont {Pierunek}},
  \ and\ \bibinfo {author} {\bibfnamefont {B.}~\bibnamefont {Idzikowski}},\
  }\href {\doibase 10.1007/s00339-014-8829-x} {\bibfield  {journal} {\bibinfo
  {journal} {Applied Physics A}\ }\textbf {\bibinfo {volume} {118}},\ \bibinfo
  {pages} {1273} (\bibinfo {year} {2015})}\BibitemShut {NoStop}%
\bibitem [{\citenamefont {Eriksson}\ \emph {et~al.}(2017)\citenamefont
  {Eriksson}, \citenamefont {Bergman}, \citenamefont {Bergqvist},\ and\
  \citenamefont {Hellsvik}}]{eriksson_atomistic_2017}%
  \BibitemOpen
  \bibfield  {author} {\bibinfo {author} {\bibfnamefont {O.}~\bibnamefont
  {Eriksson}}, \bibinfo {author} {\bibfnamefont {A.}~\bibnamefont {Bergman}},
  \bibinfo {author} {\bibfnamefont {L.}~\bibnamefont {Bergqvist}}, \ and\
  \bibinfo {author} {\bibfnamefont {J.}~\bibnamefont {Hellsvik}},\ }\href@noop
  {} {\emph {\bibinfo {title} {Atomistic {Spin} {Dynamics}: {Foundations} and
  {Applications}}}}\ (\bibinfo  {publisher} {Oxford University Press},\
  \bibinfo {year} {2017})\BibitemShut {NoStop}%
\bibitem [{\citenamefont {Gyorffy}\ \emph {et~al.}(1985)\citenamefont
  {Gyorffy}, \citenamefont {Pindor}, \citenamefont {Staunton}, \citenamefont
  {Stocks},\ and\ \citenamefont {Winter}}]{gyorffy_first-principles_1985}%
  \BibitemOpen
  \bibfield  {author} {\bibinfo {author} {\bibfnamefont {B.~L.}\ \bibnamefont
  {Gyorffy}}, \bibinfo {author} {\bibfnamefont {A.~J.}\ \bibnamefont {Pindor}},
  \bibinfo {author} {\bibfnamefont {J.}~\bibnamefont {Staunton}}, \bibinfo
  {author} {\bibfnamefont {G.~M.}\ \bibnamefont {Stocks}}, \ and\ \bibinfo
  {author} {\bibfnamefont {H.}~\bibnamefont {Winter}},\ }\href {\doibase
  10.1088/0305-4608/15/6/018} {\bibfield  {journal} {\bibinfo  {journal}
  {Journal of Physics F: Metal Physics}\ }\textbf {\bibinfo {volume} {15}},\
  \bibinfo {pages} {1337} (\bibinfo {year} {1985})}\BibitemShut {NoStop}%
\bibitem [{\citenamefont {Bergqvist}\ and\ \citenamefont
  {Dederichs}(2007)}]{bergqvist_theoretical_2007}%
  \BibitemOpen
  \bibfield  {author} {\bibinfo {author} {\bibfnamefont {L.}~\bibnamefont
  {Bergqvist}}\ and\ \bibinfo {author} {\bibfnamefont {P.~H.}\ \bibnamefont
  {Dederichs}},\ }\href {\doibase 10.1088/0953-8984/19/21/216220} {\bibfield
  {journal} {\bibinfo  {journal} {Journal of Physics: Condensed Matter}\
  }\textbf {\bibinfo {volume} {19}},\ \bibinfo {pages} {216220} (\bibinfo
  {year} {2007})}\BibitemShut {NoStop}%
\bibitem [{\citenamefont {Soven}(1967)}]{soven_coherent-potential_1967}%
  \BibitemOpen
  \bibfield  {author} {\bibinfo {author} {\bibfnamefont {P.}~\bibnamefont
  {Soven}},\ }\href
  {http://journals.aps.org/pr/abstract/10.1103/PhysRev.156.809} {\bibfield
  {journal} {\bibinfo  {journal} {Physical Review}\ }\textbf {\bibinfo {volume}
  {156}},\ \bibinfo {pages} {809} (\bibinfo {year} {1967})}\BibitemShut
  {NoStop}%
\bibitem [{\citenamefont {Skubic}\ \emph {et~al.}(2008)\citenamefont {Skubic},
  \citenamefont {Hellsvik}, \citenamefont {Nordstr{\"o}m},\ and\ \citenamefont
  {Eriksson}}]{skubic_method_2008}%
  \BibitemOpen
  \bibfield  {author} {\bibinfo {author} {\bibfnamefont {B.}~\bibnamefont
  {Skubic}}, \bibinfo {author} {\bibfnamefont {J.}~\bibnamefont {Hellsvik}},
  \bibinfo {author} {\bibfnamefont {L.}~\bibnamefont {Nordstr{\"o}m}}, \ and\
  \bibinfo {author} {\bibfnamefont {O.}~\bibnamefont {Eriksson}},\ }\href
  {\doibase 10.1088/0953-8984/20/31/315203} {\bibfield  {journal} {\bibinfo
  {journal} {Journal of Physics: Condensed Matter}\ }\textbf {\bibinfo {volume}
  {20}},\ \bibinfo {pages} {315203} (\bibinfo {year} {2008})}\BibitemShut
  {NoStop}%
\bibitem [{\citenamefont {Ebert~et al.}()}]{ebert_et_al._munich_nodate}%
  \BibitemOpen
  \bibfield  {author} {\bibinfo {author} {\bibfnamefont {H.}~\bibnamefont
  {Ebert~et al.}},\ }\href {http://ebert.cup. uni-muenchen.de/SPRKKR} {\enquote
  {\bibinfo {title} {The {Munich} {SPR}-{KKR} {Package}, {Version} 6.3
  http://ebert.cup. uni-muenchen.de/{SPRKKR}},}\ }\BibitemShut {NoStop}%
\bibitem [{\citenamefont {Ebert}\ \emph {et~al.}(2011)\citenamefont {Ebert},
  \citenamefont {K{\"o}dderitzsch},\ and\ \citenamefont
  {Min{\'a}r}}]{ebert_calculating_2011}%
  \BibitemOpen
  \bibfield  {author} {\bibinfo {author} {\bibfnamefont {H.}~\bibnamefont
  {Ebert}}, \bibinfo {author} {\bibfnamefont {D.}~\bibnamefont
  {K{\"o}dderitzsch}}, \ and\ \bibinfo {author} {\bibfnamefont
  {J.}~\bibnamefont {Min{\'a}r}},\ }\href {\doibase
  10.1088/0034-4885/74/9/096501} {\bibfield  {journal} {\bibinfo  {journal}
  {Reports on Progress in Physics}\ }\textbf {\bibinfo {volume} {74}},\
  \bibinfo {pages} {096501} (\bibinfo {year} {2011})}\BibitemShut {NoStop}%
\bibitem [{\citenamefont {Liechtenstein}\ \emph {et~al.}(1984)\citenamefont
  {Liechtenstein}, \citenamefont {Katsnelson},\ and\ \citenamefont
  {Gubanov}}]{liechtenstein_exchange_1984}%
  \BibitemOpen
  \bibfield  {author} {\bibinfo {author} {\bibfnamefont {A.~I.}\ \bibnamefont
  {Liechtenstein}}, \bibinfo {author} {\bibfnamefont {M.~I.}\ \bibnamefont
  {Katsnelson}}, \ and\ \bibinfo {author} {\bibfnamefont {V.~A.}\ \bibnamefont
  {Gubanov}},\ }\href {\doibase 10.1088/0305-4608/14/7/007} {\bibfield
  {journal} {\bibinfo  {journal} {Journal of Physics F: Metal Physics}\
  }\textbf {\bibinfo {volume} {14}},\ \bibinfo {pages} {L125} (\bibinfo {year}
  {1984})}\BibitemShut {NoStop}%
\bibitem [{\citenamefont {Momma}\ and\ \citenamefont
  {Izumi}(2011)}]{momma_vesta_2011}%
  \BibitemOpen
  \bibfield  {author} {\bibinfo {author} {\bibfnamefont {K.}~\bibnamefont
  {Momma}}\ and\ \bibinfo {author} {\bibfnamefont {F.}~\bibnamefont {Izumi}},\
  }\href {\doibase 10.1107/S0021889811038970} {\bibfield  {journal} {\bibinfo
  {journal} {Journal of Applied Crystallography}\ }\textbf {\bibinfo {volume}
  {44}},\ \bibinfo {pages} {1272} (\bibinfo {year} {2011})}\BibitemShut
  {NoStop}%
\bibitem [{\citenamefont {Muraoka}\ \emph {et~al.}(1979)\citenamefont
  {Muraoka}, \citenamefont {Shiga},\ and\ \citenamefont
  {Nakamura}}]{muraoka_magnetic_1979}%
  \BibitemOpen
  \bibfield  {author} {\bibinfo {author} {\bibfnamefont {Y.}~\bibnamefont
  {Muraoka}}, \bibinfo {author} {\bibfnamefont {M.}~\bibnamefont {Shiga}}, \
  and\ \bibinfo {author} {\bibfnamefont {Y.}~\bibnamefont {Nakamura}},\
  }\href@noop {} {\bibfield  {journal} {\bibinfo  {journal} {Journal of Physics
  F: Metal Physics}\ }\textbf {\bibinfo {volume} {9}},\ \bibinfo {pages} {1889}
  (\bibinfo {year} {1979})}\BibitemShut {NoStop}%
\bibitem [{\citenamefont {Koepernik}\ \emph {et~al.}(1997)\citenamefont
  {Koepernik}, \citenamefont {Velick{\'y}}, \citenamefont {Hayn},\ and\
  \citenamefont {Eschrig}}]{koepernik_self-consistent_1997}%
  \BibitemOpen
  \bibfield  {author} {\bibinfo {author} {\bibfnamefont {K.}~\bibnamefont
  {Koepernik}}, \bibinfo {author} {\bibfnamefont {B.}~\bibnamefont
  {Velick{\'y}}}, \bibinfo {author} {\bibfnamefont {R.}~\bibnamefont {Hayn}}, \
  and\ \bibinfo {author} {\bibfnamefont {H.}~\bibnamefont {Eschrig}},\ }\href
  {http://journals.aps.org/prb/abstract/10.1103/PhysRevB.55.5717} {\bibfield
  {journal} {\bibinfo  {journal} {Physical Review B}\ }\textbf {\bibinfo
  {volume} {55}},\ \bibinfo {pages} {5717} (\bibinfo {year}
  {1997})}\BibitemShut {NoStop}%
\bibitem [{\citenamefont {Koepernik}\ and\ \citenamefont
  {Eschrig}(1999)}]{koepernik_full-potential_1999}%
  \BibitemOpen
  \bibfield  {author} {\bibinfo {author} {\bibfnamefont {K.}~\bibnamefont
  {Koepernik}}\ and\ \bibinfo {author} {\bibfnamefont {H.}~\bibnamefont
  {Eschrig}},\ }\href
  {http://journals.aps.org/prb/abstract/10.1103/PhysRevB.59.1743} {\bibfield
  {journal} {\bibinfo  {journal} {Physical Review B}\ }\textbf {\bibinfo
  {volume} {59}},\ \bibinfo {pages} {1743} (\bibinfo {year}
  {1999})}\BibitemShut {NoStop}%
\bibitem [{\citenamefont {Perdew}\ and\ \citenamefont
  {Wang}(1992)}]{perdew_accurate_1992}%
  \BibitemOpen
  \bibfield  {author} {\bibinfo {author} {\bibfnamefont {J.~P.}\ \bibnamefont
  {Perdew}}\ and\ \bibinfo {author} {\bibfnamefont {Y.}~\bibnamefont {Wang}},\
  }\href {http://journals.aps.org/prb/abstract/10.1103/PhysRevB.45.13244}
  {\bibfield  {journal} {\bibinfo  {journal} {Physical Review B}\ }\textbf
  {\bibinfo {volume} {45}},\ \bibinfo {pages} {13244} (\bibinfo {year}
  {1992})}\BibitemShut {NoStop}%
\bibitem [{\citenamefont {Vosko}\ \emph {et~al.}(1980)\citenamefont {Vosko},
  \citenamefont {Wilk},\ and\ \citenamefont {Nusair}}]{vosko_accurate_1980}%
  \BibitemOpen
  \bibfield  {author} {\bibinfo {author} {\bibfnamefont {S.~H.}\ \bibnamefont
  {Vosko}}, \bibinfo {author} {\bibfnamefont {L.}~\bibnamefont {Wilk}}, \ and\
  \bibinfo {author} {\bibfnamefont {M.}~\bibnamefont {Nusair}},\ }\href
  {\doibase 10.1139/p80-159} {\bibfield  {journal} {\bibinfo  {journal}
  {Canadian Journal of Physics}\ }\textbf {\bibinfo {volume} {58}},\ \bibinfo
  {pages} {1200} (\bibinfo {year} {1980})}\BibitemShut {NoStop}%
\bibitem [{\citenamefont {Kilcoyne}(2000)}]{kilcoyne_evolution_2000}%
  \BibitemOpen
  \bibfield  {author} {\bibinfo {author} {\bibfnamefont {S.~H.}\ \bibnamefont
  {Kilcoyne}},\ }\href {\doibase 10.1016/S0921-4526(99)01731-7} {\bibfield
  {journal} {\bibinfo  {journal} {Physica B: Condensed Matter}\ }\textbf
  {\bibinfo {volume} {276{\textendash}278}},\ \bibinfo {pages} {660} (\bibinfo
  {year} {2000})}\BibitemShut {NoStop}%
\bibitem [{\citenamefont {Ke}\ \emph {et~al.}(2013)\citenamefont {Ke},
  \citenamefont {Belashchenko}, \citenamefont {van Schilfgaarde}, \citenamefont
  {Kotani},\ and\ \citenamefont {Antropov}}]{ke_effects_2013}%
  \BibitemOpen
  \bibfield  {author} {\bibinfo {author} {\bibfnamefont {L.}~\bibnamefont
  {Ke}}, \bibinfo {author} {\bibfnamefont {K.~D.}\ \bibnamefont
  {Belashchenko}}, \bibinfo {author} {\bibfnamefont {M.}~\bibnamefont {van
  Schilfgaarde}}, \bibinfo {author} {\bibfnamefont {T.}~\bibnamefont {Kotani}},
  \ and\ \bibinfo {author} {\bibfnamefont {V.~P.}\ \bibnamefont {Antropov}},\
  }\href {\doibase 10.1103/PhysRevB.88.024404} {\bibfield  {journal} {\bibinfo
  {journal} {Physical Review B}\ }\textbf {\bibinfo {volume} {88}} (\bibinfo
  {year} {2013}),\ 10.1103/PhysRevB.88.024404}\BibitemShut {NoStop}%
\bibitem [{\citenamefont {Hedlund}\ \emph {et~al.}(2017)\citenamefont
  {Hedlund}, \citenamefont {Cedervall}, \citenamefont {Edstr{\"o}m},
  \citenamefont {Werwi{\'n}ski}, \citenamefont {Kontos}, \citenamefont
  {Eriksson}, \citenamefont {Rusz}, \citenamefont {Svedlindh}, \citenamefont
  {Sahlberg},\ and\ \citenamefont {Gunnarsson}}]{hedlund_magnetic_2017}%
  \BibitemOpen
  \bibfield  {author} {\bibinfo {author} {\bibfnamefont {D.}~\bibnamefont
  {Hedlund}}, \bibinfo {author} {\bibfnamefont {J.}~\bibnamefont {Cedervall}},
  \bibinfo {author} {\bibfnamefont {A.}~\bibnamefont {Edstr{\"o}m}}, \bibinfo
  {author} {\bibfnamefont {M.}~\bibnamefont {Werwi{\'n}ski}}, \bibinfo {author}
  {\bibfnamefont {S.}~\bibnamefont {Kontos}}, \bibinfo {author} {\bibfnamefont
  {O.}~\bibnamefont {Eriksson}}, \bibinfo {author} {\bibfnamefont
  {J.}~\bibnamefont {Rusz}}, \bibinfo {author} {\bibfnamefont {P.}~\bibnamefont
  {Svedlindh}}, \bibinfo {author} {\bibfnamefont {M.}~\bibnamefont {Sahlberg}},
  \ and\ \bibinfo {author} {\bibfnamefont {K.}~\bibnamefont {Gunnarsson}},\
  }\href {\doibase 10.1103/PhysRevB.96.094433} {\bibfield  {journal} {\bibinfo
  {journal} {Physical Review B}\ }\textbf {\bibinfo {volume} {96}} (\bibinfo
  {year} {2017}),\ 10.1103/PhysRevB.96.094433}\BibitemShut {NoStop}%
\bibitem [{\citenamefont {Rusz}\ \emph {et~al.}(2006)\citenamefont {Rusz},
  \citenamefont {Bergqvist}, \citenamefont {Kudrnovsk{\'y}},\ and\
  \citenamefont {Turek}}]{rusz_exchange_2006}%
  \BibitemOpen
  \bibfield  {author} {\bibinfo {author} {\bibfnamefont {J.}~\bibnamefont
  {Rusz}}, \bibinfo {author} {\bibfnamefont {L.}~\bibnamefont {Bergqvist}},
  \bibinfo {author} {\bibfnamefont {J.}~\bibnamefont {Kudrnovsk{\'y}}}, \ and\
  \bibinfo {author} {\bibfnamefont {I.}~\bibnamefont {Turek}},\ }\href
  {\doibase 10.1103/PhysRevB.73.214412} {\bibfield  {journal} {\bibinfo
  {journal} {Physical Review B}\ }\textbf {\bibinfo {volume} {73}} (\bibinfo
  {year} {2006}),\ 10.1103/PhysRevB.73.214412}\BibitemShut {NoStop}%
\bibitem [{\citenamefont {Khmelevskyi}\ \emph {et~al.}(2005)\citenamefont
  {Khmelevskyi}, \citenamefont {Mohn}, \citenamefont {Redinger},\ and\
  \citenamefont {Weinert}}]{khmelevskyi_magnetism_2005}%
  \BibitemOpen
  \bibfield  {author} {\bibinfo {author} {\bibfnamefont {S.}~\bibnamefont
  {Khmelevskyi}}, \bibinfo {author} {\bibfnamefont {P.}~\bibnamefont {Mohn}},
  \bibinfo {author} {\bibfnamefont {J.}~\bibnamefont {Redinger}}, \ and\
  \bibinfo {author} {\bibfnamefont {M.}~\bibnamefont {Weinert}},\ }\href
  {\doibase 10.1103/PhysRevLett.94.146403} {\bibfield  {journal} {\bibinfo
  {journal} {Physical Review Letters}\ }\textbf {\bibinfo {volume} {94}}
  (\bibinfo {year} {2005}),\ 10.1103/PhysRevLett.94.146403}\BibitemShut
  {NoStop}%
\bibitem [{\citenamefont {Edstr{\"o}m}\ \emph {et~al.}(2015)\citenamefont
  {Edstr{\"o}m}, \citenamefont {Werwi{\'n}ski}, \citenamefont {Iu{\c s}an},
  \citenamefont {Rusz}, \citenamefont {Eriksson}, \citenamefont {Skokov},
  \citenamefont {Radulov}, \citenamefont {Ener}, \citenamefont {Kuz'min},
  \citenamefont {Hong}, \citenamefont {Fries}, \citenamefont {Karpenkov},
  \citenamefont {Gutfleisch}, \citenamefont {Toson},\ and\ \citenamefont
  {Fidler}}]{edstrom_magnetic_2015}%
  \BibitemOpen
  \bibfield  {author} {\bibinfo {author} {\bibfnamefont {A.}~\bibnamefont
  {Edstr{\"o}m}}, \bibinfo {author} {\bibfnamefont {M.}~\bibnamefont
  {Werwi{\'n}ski}}, \bibinfo {author} {\bibfnamefont {D.}~\bibnamefont {Iu{\c
  s}an}}, \bibinfo {author} {\bibfnamefont {J.}~\bibnamefont {Rusz}}, \bibinfo
  {author} {\bibfnamefont {O.}~\bibnamefont {Eriksson}}, \bibinfo {author}
  {\bibfnamefont {K.~P.}\ \bibnamefont {Skokov}}, \bibinfo {author}
  {\bibfnamefont {I.~A.}\ \bibnamefont {Radulov}}, \bibinfo {author}
  {\bibfnamefont {S.}~\bibnamefont {Ener}}, \bibinfo {author} {\bibfnamefont
  {M.~D.}\ \bibnamefont {Kuz'min}}, \bibinfo {author} {\bibfnamefont
  {J.}~\bibnamefont {Hong}}, \bibinfo {author} {\bibfnamefont {M.}~\bibnamefont
  {Fries}}, \bibinfo {author} {\bibfnamefont {D.~Y.}\ \bibnamefont
  {Karpenkov}}, \bibinfo {author} {\bibfnamefont {O.}~\bibnamefont
  {Gutfleisch}}, \bibinfo {author} {\bibfnamefont {P.}~\bibnamefont {Toson}}, \
  and\ \bibinfo {author} {\bibfnamefont {J.}~\bibnamefont {Fidler}},\ }\href
  {\doibase 10.1103/PhysRevB.92.174413} {\bibfield  {journal} {\bibinfo
  {journal} {Physical Review B}\ }\textbf {\bibinfo {volume} {92}},\ \bibinfo
  {pages} {174413} (\bibinfo {year} {2015})}\BibitemShut {NoStop}%
\bibitem [{\citenamefont {Kotliar}\ \emph {et~al.}(2006)\citenamefont
  {Kotliar}, \citenamefont {Savrasov}, \citenamefont {Haule}, \citenamefont
  {Oudovenko}, \citenamefont {Parcollet},\ and\ \citenamefont
  {Marianetti}}]{kotliar_electronic_2006}%
  \BibitemOpen
  \bibfield  {author} {\bibinfo {author} {\bibfnamefont {G.}~\bibnamefont
  {Kotliar}}, \bibinfo {author} {\bibfnamefont {S.~Y.}\ \bibnamefont
  {Savrasov}}, \bibinfo {author} {\bibfnamefont {K.}~\bibnamefont {Haule}},
  \bibinfo {author} {\bibfnamefont {V.~S.}\ \bibnamefont {Oudovenko}}, \bibinfo
  {author} {\bibfnamefont {O.}~\bibnamefont {Parcollet}}, \ and\ \bibinfo
  {author} {\bibfnamefont {C.~A.}\ \bibnamefont {Marianetti}},\ }\href
  {\doibase 10.1103/RevModPhys.78.865} {\bibfield  {journal} {\bibinfo
  {journal} {Reviews of Modern Physics}\ }\textbf {\bibinfo {volume} {78}},\
  \bibinfo {pages} {865} (\bibinfo {year} {2006})}\BibitemShut {NoStop}%
\end{thebibliography}%

\end{document}